\documentclass[preprint,showpacs,amsmath,amssymb]{revtex4}
\usepackage[mathscr]{eucal}
\usepackage{graphicx}

\def\dd{\mathrm{d}}

\def\ee{{\rm e}}

\newcommand{\A}{\mathscr{A}}
\newcommand{\B}{\mathscr{B}}

\newcommand{\F}{\mathscr{F}}

\newcommand{\J}{\mathscr{J}}

\newcommand{\PP}{\mathscr{P}}

\def\la{\left\langle}
\def\ra{\right\rangle}

\providecommand\bnabla{\boldsymbol{\nabla}}
\providecommand\bcdot{\boldsymbol{\cdot}}
\newcommand{\bx}{\ensuremath {\boldsymbol {x}}}

\newcommand{\bu}{\ensuremath {\boldsymbol {u}}}

\newcommand\be{\boldsymbol{\hat e}}

\newcommand\bxi{\boldsymbol{\xi}}

\newcommand{\half}{\mbox{$\frac12$}}

\newcommand{\quarter}{\mbox{$\frac14$}}

\def\hs{\tilde{h}}

\def\gammam{\gamma_{\mathrm{max}}}

\def\sat{\eta}
\def\kappastar{\kappa_*}
\def\Ustar{U_*}
\newcommand{\da}{\mathrm{Da}}
\newcommand{\pe}{\mathrm{Pe}}

\begin{document}
%\preprint{???}
\title{Bounding biomass in the Fisher equation}

\author{Daniel~A.~Birch}
\email{dbirch@ucsd.edu}
\author{Yue-Kin Tsang}
\email{yktsang@ucsd.edu}
\author{William~R.~Young}
\email{wryoung@ucsd.edu}

\affiliation{Scripps Institution of Oceanography,
                  University of California at San Diego,\\
                  La Jolla,  CA 92093--0213, USA}

\date{\today}

\begin{abstract}
 The FKPP equation with a variable growth rate and advection by an incompressible velocity field is considered as a model for plankton dispersed by ocean currents.  If the average growth rate is negative then the model has a survival-extinction transition;  the location of this transition in the parameter space is constrained using variational arguments and delimited by simulations.  The statistical steady state reached when the system is in the survival region of parameter space is characterized by integral constraints and upper and lower bounds on the biomass and productivity that follow from variational arguments and direct inequalities.  In the limit of zero-decorrelation time the velocity field is shown to act as Fickian diffusion with an eddy diffusivity much larger than the molecular diffusivity and this allows a one-dimensional model to predict the biomass, productivity and extinction transitions. All results are illustrated with a simple growth and stirring model.
\end{abstract}

\pacs{47.70.Fw, 92.20.Jt, 87.23.Cc, 05.45.-a}

%\keywords{Fisher equation, logistic nonlinearity, population dynamics, strange eigenmode}

\maketitle

\section{Introduction}

\begin{figure*}
\begin{center}
\mbox{\includegraphics[width=.75\textwidth]{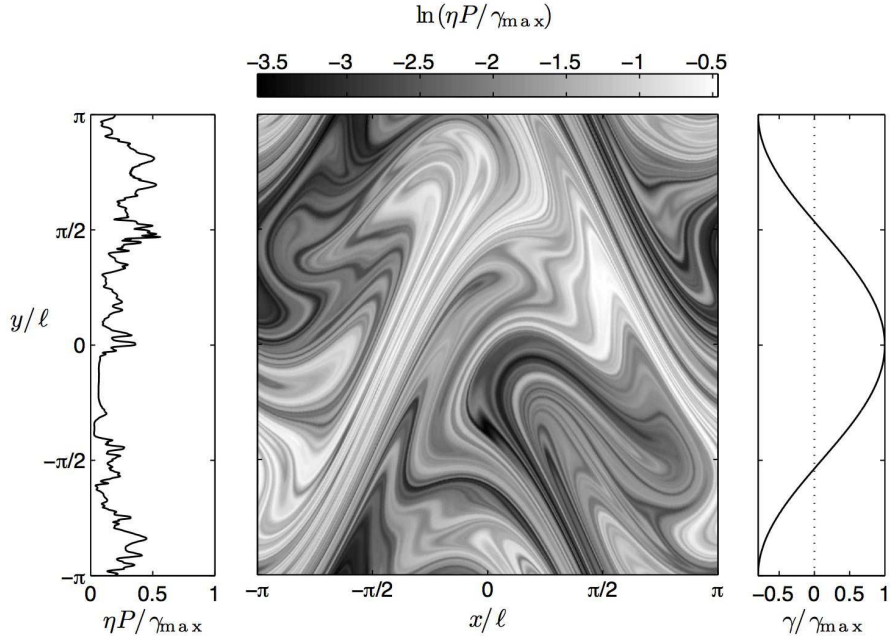}}
\end{center}
\caption{Center:  A snapshot of $\ln(P)$ from a simulation of \eqref{sle}.  The plankton concentration along the line $x = 0$ is shown on the left and the growth rate $\gamma(y)$ is on the right.  This simulation uses the model in section \textbf{\ref{modelSection}} with $\Gamma = 0.1$, $\Ustar = 10$, $m=1$, $\tau_*=0.25$ and $\kappa_* = 1\times10^{-4}$.  }
\label{adrDemoFig}
\end{figure*}

Fisher \cite{Fisher1937} and Kolmogorov, Petrovskii and Piskunov \cite{KPP1937}  introduced a partial differential equation,  now called the FKPP equation, modelling the spread via diffusion of an advantageous gene through a dispersed population.  Skellam \cite{Skellam1951} applied the FKPP equation to the change in abundance of organisms in space and time.  Oceanographic applications, particularly the dynamics of plankton populations, motivate extending the FKPP model by inclusion of   an incompressible velocity $\bu(\bx,t)$. Thus the FKPP equation considered here is
\begin{equation}
P_t+ \bu \!\bcdot\! \bnabla P=\gamma P - \sat P^2 + \kappa \nabla^2 P\, ,  
\label{sle}
\end{equation} 
where $P(\bx,t)$ is the concentration of phytoplankton.
This is the simplest model containing the four essential ingredients of advection, growth, saturation and diffusion.   Because of environmental variability, the growth rate $\gamma(\bx, t)$ may depend on both location $\bx$ and time $t$.  The small-scale diffusivity $\kappa$ and the saturation coefficient $\eta$ are taken to be positive constants.  Figure \ref{adrDemoFig}  shows a snapshot of a typical solution of (\ref{sle}).

If $\gamma$ in \eqref{sle} is a positive constant then a small initial population grows to occupy the entire domain so that ultimately  the concentration is uniform (i.e. $\lim_{t \to \infty}P = \gamma/\eta$).  The most interesting aspect of this special case is the interaction between front propagation and advection which occurs on the way to the uniform steady state \citep[e.g.][]{AudolyBP2000, NeufeldHT2002}.
On the other hand, in Figure \ref{adrDemoFig}, $\bu(\bx,t)$ continually stirs the population relative to the spatially variable growth rate and carrying capacity, resulting in a non-trivial statistically steady solution.

In oceanography a great deal of effort has gone into studying ``plankton patchiness'' \cite{APMartin2003} and the small-scale structures  produced by lateral stirring \citep[e.g.][]{NeufeldLH1999}.  Here we avoid the issue of defining a patch or patchiness, and we also largely avoid considerations of the small-scale structure evident in Figure \ref{adrDemoFig}. Instead we consider a more basic question:  can we predict or constrain the total amount of plankton in solutions of \eqref{sle}?  To pursue this goal we develop upper and lower bounds on the plankton biomass which depend only on the gross properties of the growth rate and stirring.   In addition to bounds on the biomass, we also develop bounds on the productivity, which may be related to the variance of the concentration. These bounds are obtained using mathematical techniques  with parallel applications to the Navier-Stokes equation and the passive scalar problem \citep[e.g.][]{DoeringC1994, PlastingY2006, DoeringT2006}.

In Section \textbf{\ref{modelSection}} we describe the model growth rate and velocity used  to illustrate our general results and we make a  comparison between an inert scalar and the reactive tracer $P$ in \eqref{sle}.  The survival-extinction transition is discussed in Section \textbf{\ref{extSection}}.  In Sections \textbf{\ref{boundsSection}} and \textbf{\ref{jointSection}} we develop inequalities which constrain the biomass and productivity.  Section \textbf{\ref{conclusions}} contains the conclusions and discussion. Some mathematical details are contained in three appendices.

\section{An illustrative growth rate and velocity field \label{modelSection}}

\begin{figure*}
\begin{center}
\mbox{\includegraphics[width=.75\textwidth]{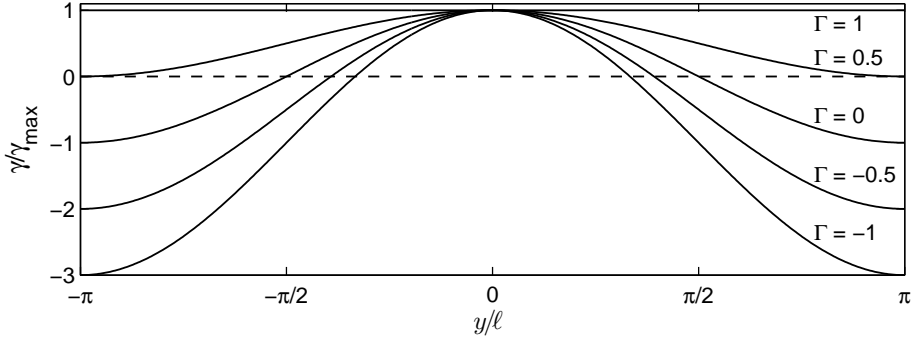}}
\end{center}
\caption{The model growth rate \eqref{sinugamma} for various $\Gamma$.}
\label{gammaFig}
\end{figure*}

Our main results will apply to a variety of space- and time-dependent growth rates and flows.  However, for the sake of illustration, we will present examples using a model defined on a doubly periodic square domain, with $x,y \in [-\pi\ell, \pi\ell)$, and the growth rate given by the ``sinusoidal'' model:
\begin{equation}
\gamma(y) = \gammam\left[ \Gamma + (1-\Gamma)\cos{\left(y/\ell\right)} \right] \, , \quad y \in [-\pi\ell, \pi\ell) \, .
\label{sinugamma}
\end{equation}
$\Gamma\in(-\infty, 1]$ is non-dimensional and controls both the average and the spatial structure of the growth rate as shown in Figure \ref{gammaFig}.

Stirring is provided by a popular model of a random two-dimensional velocity field \citep[e.g][]{BJBayly1992, STF, RTPierrehumbert1994, RTPierrehumbert2000, AntonsenFOG-L1996}.   The velocity alternates between $v=0$ and  $u=0$:
\begin{equation}
\bu(\bx,t)= \begin{cases}  \sqrt{2}U \left(\cos( k_m y +\phi_x)\, , \, 0\right)\, , & \text{for $n \tau \leq t < (n+1/2) \tau$}\, ;  \\
\sqrt{2} U \left(0\, , \, \cos(k_m x+\phi_y)\right)\, , & \text{for  $(n+1/2) \tau \leq t < (n+1)\tau$}\, .
\end{cases}
\label{renwave1}
\end{equation}
Above, $k_m=m/\ell$ where $m$ is an integer controlling the scale separation between the domain scale $\ell$ and the velocity field. The phases $\phi_x$ and $\phi_y$ are randomly chosen each period with uniform density on $(0,2 \pi)$.

The average squared velocity components of the stirring model \eqref{renwave1} are $\la u^2\ra = \la v^2 \ra = \half{}U^2$, and the flow is homogeneous and isotropic in the sense that
\begin{equation}
\la \bu_i \bu_j \ra = \half U^2 \delta_{i j}\, .
\label{shif}
\end{equation} 
The angled braces in \eqref{shif} indicate a space-time average and are explicitly defined in \eqref{eq1}. Since the flow is isotropic, the single-particle eddy diffusivity may be found by using the relationship 
\begin{equation}
2D\tau = \la {(\Delta x)}^2 \ra = \la {(\Delta y)}^2 \ra \, ,
\end{equation}
where $\Delta x$ and $\Delta y$ are the $x$- and $y$-displacements of a particle during the time interval $t = 0$ to $\tau$.  Thus the eddy diffusivity of \eqref{renwave1} is
\begin{equation}
D = \frac{U^2\tau}{8} \, .
\label{renwave2}
\end{equation}

\begin{figure*}
\begin{center}
\mbox{\includegraphics[width=.75\textwidth]{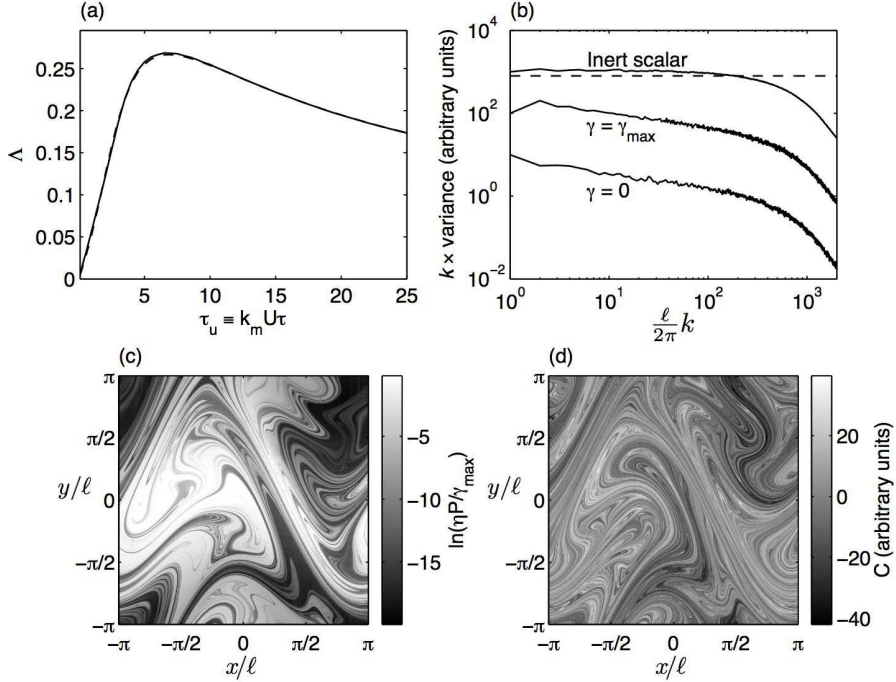}}
\end{center}
\caption{(a) The solid curve shows a Monte-Carlo estimate of the non-dimensional Lyapunov function $\Lambda(\tau_u)$ defined in \eqref{LyapDef1}; the dashed curve is the fit in \eqref{LyapDef2}.  (b) Spectra of the inert scalar $C$ in \eqref{renwave4} and of the biological tracer $P$ obtained from \eqref{sle}. The spectra are ``compensated'' by multiplying the variance spectrum by the wave number so that a $k^{-1}$ Batchelor spectrum appears flat. The plankton are growing according to the sinusoidal growth rate in \eqref{sinugamma} with $\Gamma=0$. The $P$-spectra are line spectra at $y/\ell=0$, where $\gamma = \gammam$, and at $y/\ell=\pi/2$, where $\gamma=0$. $C$ is well described by the Batchelor spectrum, while $P$ has a steeper spectral slope. Panels (c) and (d) show snapshots of the $P$ and $C$ respectively; note in panel (c) the contour interval is logarithmic (i.e., panel (c) shows $Z \equiv \ln(P)$).  Both simulations use the parameters $\kappa_*=10^{-7}$, $U_*=1$, $\tau_*=2.2214$ and $m=1$. }
\label{LyapFig}
\end{figure*}

In addition to the eddy diffusion of individual particles, we also consider the stretching and compression of an infinitesimal material line element $\bxi$. The length of the  element, $|\bxi| = \sqrt{\bxi \! \bcdot \! \bxi}$, grows exponentially at a rate  estimated by
\begin{equation}
\lambda = \lim_{t \to \infty} t^{-1} \mathbb{E}\left[ \ln | \bxi|  \right] \, ,
\label{LyapDef}
\end{equation}
where $\lambda$ is the Lyapunov exponent of the flow in \eqref{renwave1} and  $\mathbb{E}$ denotes the expectation obtained by averaging over an ensemble of material elements. Dimensional considerations show that for \eqref{renwave1} the Lyapunov exponent $\lambda$ has the form 
\begin{equation}
\lambda = Uk_m \Lambda (\tau_u)\, , \qquad \tau_u \equiv U k_m \tau\, .
\label{LyapDef1}
\end{equation}
The non-dimensional parameter $\tau_u$ is the ratio of the de-correlation time $\tau$ to the shear $\left(Uk_m\right)^{-1}$.
An estimate of the the non-dimensional function $\Lambda(\tau_u)$, obtained by the Monte-Carlo method summarized in appendix \textbf{\ref{LyapExpAppen}}, is shown as the solid curve in Figure \ref{LyapFig}(a). The dashed curve in Figure \ref{LyapFig}(a) is the approximation:
\begin{equation}
\Lambda(\tau_u) \approx \frac{\ln\left(1+ \tau_u^2/10 + \tau_u^4/67\right)}{2 \tau_u}\, .
\label{LyapDef2}
\end{equation}
This particular functional form is suggested by analytic solution of  closely related problems \cite{YoungGFD1999}.

In addition to $m$, $\Gamma$ and $\tau_u$, the model is controlled by two more parameters: the P\'eclet and Damk\"ohler numbers.  Scaling length with $\ell$ and time with $\gammam^{-1}$,  the P\'eclet number emerges as the ratio of the diffusive time scale $\ell^2/\kappa$ to the advective time scale $\ell/U$:
$
\pe \equiv {\ell U}/{\kappa} 
$.
The Damk\"ohler number is the ratio of the advective time scale $\ell/U$ to the reactive (or biological) time scale $\gammam^{-1}$:
$
 \da \equiv {\ell \gammam }/{U}
$.
For convenience, we use the equivalent parameters
\begin{equation}
\kappastar \equiv \da^{-1}\pe^{-1} = \frac{\kappa}{\ell^2 \gammam} \, , \qquad \Ustar \equiv \da^{-1} = \frac{U}{\ell\gammam}\, .
\end{equation}
The non-dimensional renovation cycle length is denoted $\tau_*$ and may be calculated from the other non-dimensional parameters:
\begin{equation}
\tau_* \equiv \gammam\tau = \frac{\tau_u}{m\Ustar} \, .
\end{equation}
Finally, the saturation constant $\eta$ may be completely removed from all equations by scaling $P$ with $\gammam/\eta$.

Using the split-step lattice method of Pierrehumbert \cite{RTPierrehumbert2000}, one can  efficiently solve both  \eqref{sle} and the forced inert passive scalar (inert scalar from now on) equation
\begin{equation}
C_t + \bu\!\bcdot\! \bnabla C = \kappa \nabla^2C + \cos k_1 y\, ,
\label{renwave4}
\end{equation}
with the velocity field in \eqref{renwave1}.
Snapshots of simulations with $m=1$ in \eqref{renwave1} and a resolution of $4096 \times 4096$ are shown in the bottom panels of Figure \ref{LyapFig}. The inert scalar $C$ in panel (d) has a classic $k^{-1}$-spectrum \cite{GKBatchelor1959} as shown in panel (b). With the sinusoidal growth rate in \eqref{sinugamma}, the statistics of the biological tracer $P$ are spatially inhomogeneous and so in Figure \ref{LyapFig}(b) we show one-dimensional $P$-spectra obtained along the lines at which $\gamma=\gammam$ and $\gamma=0$.  The $P$-spectra have slopes of $-1.26$ and $-1.40$ for the transects with $\gamma = \gammam$ and $\gamma = 0$, respectively.  The inert scalar's spectral slope is $-1.01$.  All slopes were calculated over the range of wave numbers $k = 1$ to $32$.

Deviations of the $P$-spectra from the spectra of inert scalars in the ocean are generally interpreted to mean that biological processes such as growth and grazing are strongly affecting the plankton distribution.  Continuous sampling methods have allowed comparisons of the spectra of chlorophyll in the ocean to the spectra of physical properties since 1972 \cite{TPlatt1972}, but unfortunately there is still no consensus on how the spectra should vary \cite{APMartin2003}.  However, a recent paper \cite{HodgesR2006} presents strong evidence that stirring dominates the distribution of phytoplankton along isopycnals on intermediate scales (10 -- 100 km). An added complication in the ocean is that the ``passive'' tracer most often compared to chlorophyll is temperature, which is dynamically active.

\section{The survival--extinction transition and strange eigenfunctions \label{extSection}}

Before developing bounds on the plankton biomass and productivity, we first consider a more fundamental question: do the plankton survive at all?  Early work on this issue includes classic papers on the problem of ``critical patch size'' in models without advection \cite{KiersteadS1953, Skellam1951}.  More recent work \cite{BJBayly1992, NelsonS1998, DahmenNS2000, LinMT-OLKS2004} applies to models with advection. We follow Bayly's approach \cite{BJBayly1992} by considering flows such as \eqref{renwave1} with non-zero Lyapunov exponent  and linearizing \eqref{sle}:
\begin{equation}
P_t+ \bu \!\bcdot\! \bnabla P=\gamma P  + \kappa \nabla^2 P\, .
\label{alp}
\end{equation} 
If the initial plankton concentration is very low everywhere then the quadratic nonlinearity is negligible and \eqref{alp} governs the initial behavior of the system.

\begin{figure*}
\begin{center}
\mbox{\includegraphics[width=.75\textwidth]{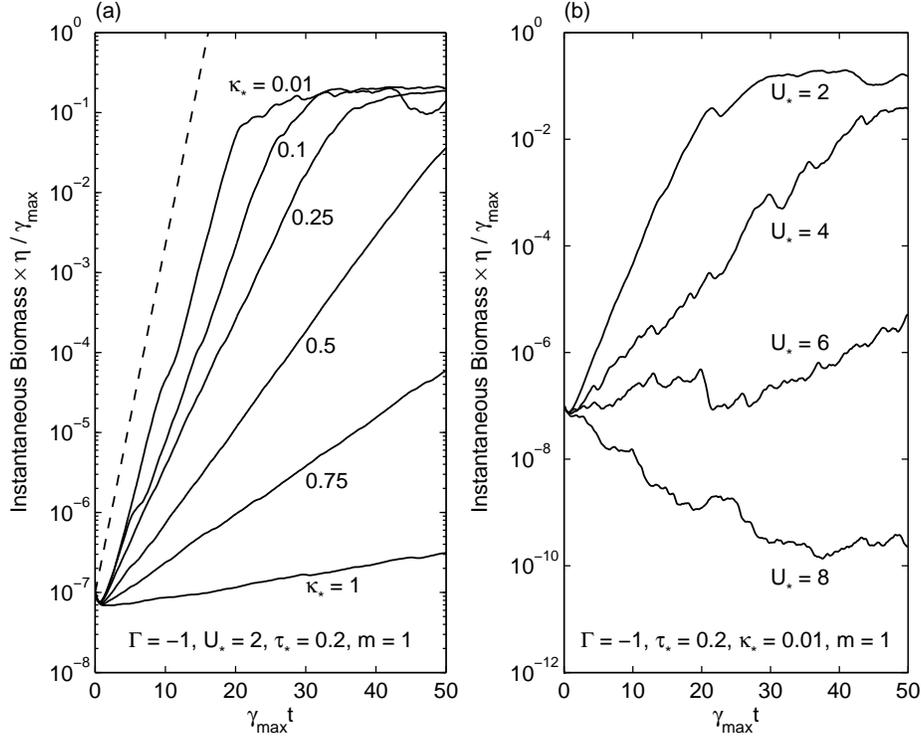}}
\end{center}
\caption{(a)  The instantaneous biomass $\bar P(t)$ shows the exponential growth of the strange eigenmode in \eqref{strange1} followed by the dynamic equilibrium when the non-linear saturation term becomes important.  The initial condition is $P(\bx,0) = 10^{-7}$ and the non-dimensional parameters are noted at the bottom of the figure.  The dashed line is the theoretical maximum $\bar{P}(t) = \bar{P}(0)\ee^{\gammam{}t}$ \cite{BJBayly1992}.  (b) Increasing $U_*$ in \eqref{renwave1} decreases $s_*$ and results in $s_*<0$ (extinction) if $U_*$ is greater than about 6.
\label{strangeGrowthFig}}
\end{figure*}

With the support of numerical simulations (see Figure \ref{strangeGrowthFig}), we assume that following a transient stage the evolution of $P$ takes the form
\begin{equation}
P(\bx,t) = \ee^{st} \hat P(\bx,t)\, .
\label{strange1}
\end{equation}
In \eqref{strange1}  the spatial distribution of $P$ is described by $\hat{P}(\bx,t)$, the statistically stationary ``strange eigenmode.''  The amplitude of the solution either grows or decays according to the sign of the ``survival exponent'' $s$ \cite{BJBayly1992, RTPierrehumbert1994}.

If $P$ is initially positive everywhere, then it will remain so and we can define
\begin{equation}
Z \equiv \ln P \, .
\label{Zdef}
\end{equation}
This is a crucial difference between $P$ and the much-studied problem of the decay of concentration anomalies of an inert scalar to zero.  Factoring $P$ into an amplitude $\ee^{st}$ and a strange eigenmode $\hat P(\bx,t)$ as in \eqref{strange1} is equivalent to writing
\begin{equation}
Z(\bx, t) = s t + \hat Z(\bx,t)\, ,
\label{strange2}
\end{equation}
where $\hat Z \equiv \ln \hat P$ is statistically stationary. In terms of $Z$ the  problem \eqref{alp} is
\begin{equation}
Z_t + \bu\bcdot\bnabla Z = \gamma(\bx,t) + \kappa \left( \nabla^2Z + {\left|\bnabla Z \right|}^2 \right) \, . \label{alp2}
\end{equation}

We begin our analysis of \eqref{alp2} by introducing a spatial average denoted by an overbar and defined by
\begin{equation}
\bar f(t)  \equiv  \frac{1}{A_{\Omega}}  \int_{\Omega}\!  f(\bx,t) \, \dd \bx  \, ,
\label{eq0}
\end{equation}
where the integral above is over the domain $\Omega$ with area $A_{\Omega}$.
For statistically stationary fields, such as $\hat Z(\bx,t)$,  we also employ a space-time average defined by 
\begin{equation}
\la f \ra \equiv  \lim_{T\to\infty}\frac{1}{TA_{\Omega}} \int_0^{T}\!\! \int_{\Omega}\!  f(\bx,t) \, \dd \bx \, \dd t  \, .
\label{eq1}
\end{equation}

Substituting \eqref{strange2} into \eqref{alp2} and spatially averaging yields
\begin{equation}
s+ \overline{\hat Z_{t}} = \bar \gamma  + \kappa \overline{ \left|\bnabla \hat Z \right|^2 }\, .
\label{alp3}
\end{equation}
Time averaging \eqref{alp3}, we obtain a fundamental connection between the survival exponent $s$, the average growth rate $\la \gamma \ra$ and $\bnabla \hat Z$:
\begin{equation}
s = \la \gamma \ra + \kappa \la \left|\bnabla \hat Z \right|^2 \ra\, .
\label{alp4}
\end{equation}
Thus if $\la \gamma \ra >0$ then  $s>0$ and the population survives (see also \cite{BJBayly1992}).  However the converse is not true: because of the final term in \eqref{alp4}, the survival exponent $s$ can be positive even if $\la \gamma \ra <0$.  In Section \ref{DiffRoleSec} we go beyond \cite{BJBayly1992} and explore survival when $\la\gamma\ra < 0$.

\subsection{An adverse environment --- the role of diffusion\label{DiffRoleSec}}

From a theoretical point of view, $\la \gamma \ra<0$ is the interesting case: the population might survive even though the average environment is adverse.  This is illustrated in Figure \ref{strangeGrowthFig}(a) where the population survives with $\Gamma = -1$.  The role of the diffusive term  $\kappa \la |\bnabla  Z |^2 \ra$ in \eqref{alp4} is quite confusing in this case and the variation of $s$ with $\kappa$ depends on the details of the flow.  In Figure \ref{strangeGrowthFig}(a) decreasing $\kappa$ increases $s$.

The limit of large diffusion and consequent extinction is straightforward: if $\kappa$ is very large then the population rapidly diffuses over the entire domain and the negative average growth rate prevails so that $s<0$. In fact, a simple perturbation expansion around $\kappa^{-1} =0$ quickly shows that
\begin{equation}
\lim_{\kappa \to \infty} s = \la \gamma \ra\, , \qquad \text{and} \qquad \lim_{\kappa \to \infty} \kappa \la \left|\bnabla \hat Z \right|^2 \ra =0\, .
\end{equation}
As always, the other limit,  $\kappa\to 0$, is potentially singular and holds the possibility that $s>0$ because the final term in \eqref{alp4} is non-zero. We investigate this possibility by consideration of some special cases and via a variational approach.

\subsection{The case $\la \gamma \ra <0$ and $\bu=0$ \label{elementary}}

\begin{figure*}
\begin{center}
\mbox{\includegraphics[width=.75\textwidth]{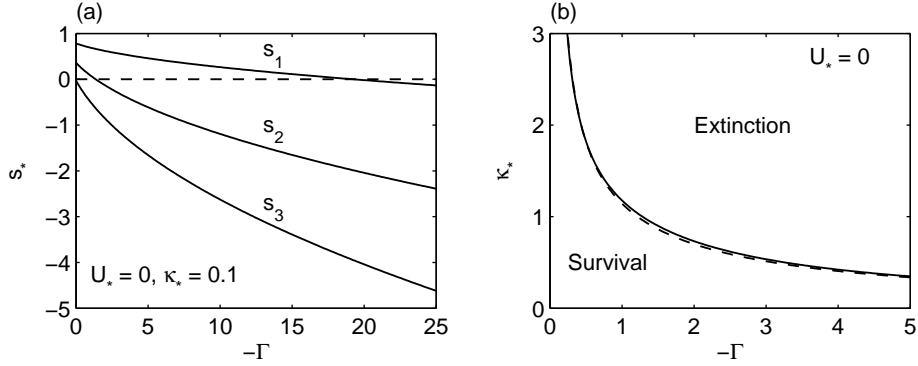}}
\end{center}
\caption{(a)  The first three eigenvalues of \eqref{zeroUeigen1}.  Extinction occurs when $s_1 < 0$, which for $\kappa_* = 0.1$ means $\Gamma\lesssim -20$.   (b) The solid curve is the $U_*=0$ extinction transition computed by numerically solving the linearized stability problem \eqref{zeroUeigen1} with $s_*=0$; the dashed curve is the approximation in \eqref{kluge1}.}
\label{zeroUFig}
\end{figure*}

As an elementary illustration of survival in an adverse environment with $\kappa \to 0$,  consider \eqref{alp} with $\bu=0$. In this case  \eqref{alp} has non-strange eigensolutions, determined by requiring $\hat P_t=0$ and substituting \eqref{strange1} into \eqref{alp}. The resulting Sturm-Liouville eigenproblem is a form of Mathieu's equation, which we express in non-dimensional variables:
  \begin{equation}
\left[\kappastar \partial_y^2 + \Gamma + (1-\Gamma) \cos y \right]\hat{P} = s_*\hat{P}  \, ,\label{zeroUeigen1}
\end{equation}
where $s_* = s/\gammam$.  Because the differential operator on the left of \eqref{zeroUeigen1} is self-adjoint, all of the eigenvalues $s_{n}$ are  real.  Figure \ref{zeroUFig}(a) shows the first three eigenvalues as a function of $-\Gamma$ for the case $\kappastar = 0.1$.

The extinction transition occurs when the largest eigenvalue $s_1$ passes through zero. Therefore, to determine the extinction transition systematically we set $s_*=0$ in \eqref{zeroUeigen1} and regard $\kappastar$ as a new eigenvalue.
The largest eigen-$\kappastar$ is the critical value of $\kappastar$, above which extinction occurs. The solid curve in Figure \ref{zeroUFig}(b) marks the boundary between survival and extinction in  $(-\Gamma,\kappa_*)$ parameter space computed in this fashion.  The dashed curve in Figure \ref{zeroUFig}(b) is an approximation to the extinction transition obtained using perturbation theory in Appendix \ref{extinctApp}:
\begin{equation}
\kappa_* \approx \frac{2+6|\Gamma |}{4|\Gamma| + 3 \Gamma^2 } \, .
\label{kluge1}
\end{equation}

Now we examine the integral constraint \eqref{alp4} in light of this example. If $\kappastar \to 0$ with $\Gamma$ fixed and negative  then the system enters the survival region in Figure \ref{zeroUFig}(b). Thus in this limit the term $\kappastar \la |\bnabla \hat Z |^2 \ra$ is both non-zero and  crucial in ensuring that $s_*>0$. Notice that since $\bu_*=0$, this singular $\kappastar \to 0$ limit does not involve gradient amplification by exponential stretching.

\begin{figure*}
\begin{center}
\mbox{\includegraphics[width=.75\textwidth]{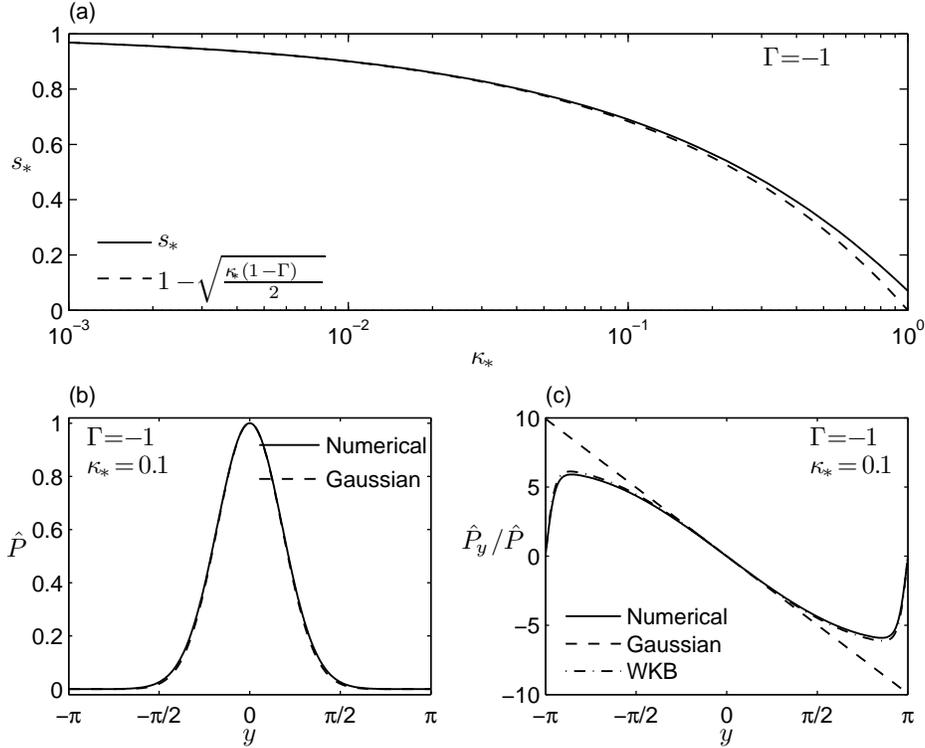}}
\end{center}
\caption{(a)  The eigenvalue $s_*$ in \eqref{zeroUeigen1} as a function of $\kappastar$.  The approximation \eqref{sApprox} improves as $\kappastar\to 0$.  (b)  The eigenmode $\hat{P}(y)$ in \eqref{zeroUeigen1} and the Gaussian approximation from Appendix \ref{extinctApp}.  The WKB approximation \eqref{PhatWKB} has a singularity near the origin and is not shown.  (c)  Three solutions for $\hat{P}_y/\hat{P}$.  Both the Gaussian and WKB approximations are good near the origin, but the Gaussian approximation fails in the region where $\gamma(y) - s <0$ and is therefore not useful for evaluating the average in \eqref{alp4}.}
\label{sAsKappaToZeroFig}
\end{figure*}

Instead, we understand the $\kappastar\to 0$ limit by finding an approximation to $\hat{P}$ and directly evaluating $\kappastar \la |\bnabla \hat Z |^2 \ra$.  The first step is calculating the largest eigenvalue of \eqref{zeroUeigen1} using the results in Appendix \ref{extinctApp} with $K = {\kappastar}/{(1-\Gamma)}$ and $E = (s_* - \Gamma)/(1-\Gamma)$ in \eqref{approx4}:
\begin{equation}
s_* \approx 1 - \sqrt{\frac{\kappastar(1-\Gamma)}{2}} \, .
\label{sApprox}
\end{equation}
As $\kappa_* \to 0$ the growth rate of the mode approaches the maximum of $\gamma$, namely $s_* \to 1$ \cite{BJBayly1992}.
Figure \ref{sAsKappaToZeroFig}(a) shows that \eqref{sApprox} is a very good approximation to $s_*$ over a wide range of $\kappastar$ and improves as $\kappastar\to 0$.  The Gaussian approximation of Appendix \ref{extinctApp}, used to obtain \eqref{sApprox}, is shown in Figure \ref{sAsKappaToZeroFig}(b) and is an excellent approximation to $\hat{P}$ in the region where the eigenmode is concentrated.  However, to reconcile $s_* \to 1$ with  the integral constraint \eqref{alp4} we need an approximation which is valid where $\hat P_y/\hat P$ is large; ironically, this is the region where $\gamma(y) -s <0$ and  $\hat P$ is very small. We can use the method of Wentzel, Kramers and Brillouin (WKB hereafter) to obtain the required approximation by assuming that $s_*\approx1$ and re-casting \eqref{zeroUeigen1} in Schr\"odinger form \cite{BenderOrszag}:
\begin{equation}
\kappastar{}\hat{P}_{yy} = R(y)\hat{P} \, , \quad R\equiv (1 - \Gamma)( 1-\cos y ) \, .
\label{schrodinger}
\end{equation}
The WKB solution to \eqref{schrodinger} which  is symmetric about $y=\pi$ is
\begin{equation}
\hat{P}_{\mathrm{WKB}} = C{R^{-1/4}(y)}\cosh\left( \kappa_*^{-1/2}\int_y^\pi \!\!\sqrt{R(y')} \,\dd y' \right) \, .
\label{PhatWKB}
\end{equation}
Now we can evaluate the term $\kappastar \la |\bnabla \hat Z |^2 \ra$ in \eqref{alp4} and find that it is independent of $\kappastar$:
\begin{eqnarray}
\kappastar \la |\bnabla \hat Z |^2 \ra & = & \kappastar \la \left(  \frac{\hat{P}_y}{\hat{P}} \right)^2 \ra \, , \nonumber \\
&\approx& \frac{1}{\pi} \!\!\int_0^\pi \!\!R(y') \tanh^2\left(\kappa_*^{-1/2}\!\!\int_y^{\pi}\!\!\sqrt{R(y'')}  \,\dd y''\right) \, \dd y' + O(\kappastar^{1/2})\, ,  \nonumber \\ 
&\approx& 1 - \Gamma + O(\kappastar^{1/2})\, ,
\label{rec}
\end{eqnarray}
where the $\tanh^2$ can be  replaced  by $1$ to evaluate the integral with errors of order $\kappa_*^{1/2}$. Equation \eqref{rec} reconciles \eqref{sApprox} with \eqref{alp4} in the singular limit $\kappastar\to 0$. Figure \ref{sAsKappaToZeroFig}(c) shows that \eqref{PhatWKB} provides good approximation to $\hat{P}_y/\hat{P}$ over the entire domain.

\subsection{The survival--extinction transition in the limit of rapid decorrelation \label{decorrelation}}

\begin{figure*}
\begin{center}
\mbox{\includegraphics[width=.75\textwidth]{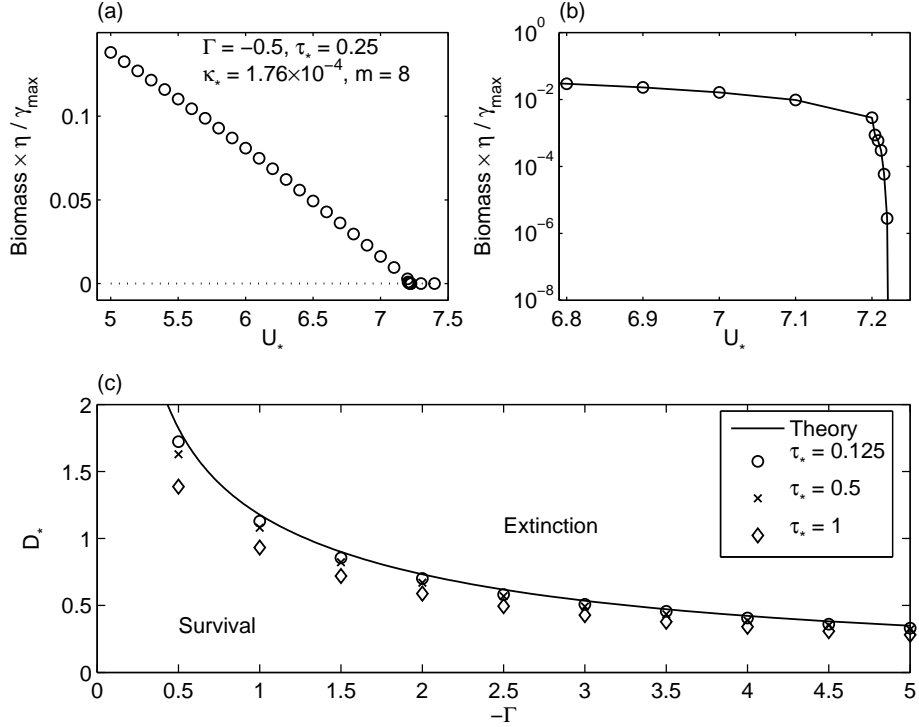}}
\end{center}
\caption{(a)  The biomass $\la P \ra$ as a function of $U_*$ as determined by Monte Carlo simulation.  The extinction transition is at around $U_*= 7.2$.  (b) An expanded view of the data in panel (a). (c) Summary of a suite of simulations in which $D_* = U^2_* \tau_*/8$ is fixed as $\tau_*$ is varied. The survival-extinction transition is located via repeated simulations with varying $U_*$, as in panels (a) and (b). As $\tau_*$ is decreased the survival-extinction transition approaches the solid curve previously shown in Figure \ref{zeroUFig}(b) with $D_*$ playing the role of $\kappa_*$.  The velocity field is \eqref{renwave1} with $m = 8$ to ensure scale separation.}
\label{YKdata}
\end{figure*}

In the  limit of a rapidly decorrelating velocity the results of the previous section can be adapted to make another quantitative prediction of the survival-extinction transition. This rapid-decorrelation  limit is achieved by taking $\tau_* \to 0$ and $U_* \to \infty$ so that the non-dimensional version of the eddy diffusivity in \eqref{renwave2}, namely $D_*\equiv U_*^2\tau_*/8$, is fixed. Provided that
\begin{equation}
D_* \gg \kappa_*\, ,
\end{equation}
and that there is scale separation between the velocity field and the domain
\begin{equation}
m \gg 1,
\end{equation}
it is plausible that the earlier eigensolution \eqref{kluge1}, with $\kappa_*$ replaced by $D_*$ in \eqref{zeroUeigen1}, can be used to locate the extinction--survival transition in the $(-\Gamma, D_*)$ parameter plane. Numerical simulations, summarized in Figure \ref{YKdata}, show that this eddy-diffusion closure works well.

\subsection{Prohibition of extinction: a lower bound on $s$ \label{prohibition}}

Aside from the rapid-decorrelation limit we do not have a simple means of determining the  location of the extinction transition in the parameter space.  However, in this section we use a simple variational method to locate a region of the $(\kappa, \la\bu^2\ra)$-plane where extinction is impossible, even if the average growth rate is negative. The region we find is contained within the (possibly larger) actual survival region. In other words, we obtain a sufficient condition for survival which applies to any incompressible velocity field without restriction to rapid decorrelation or scale separation.

We begin by  substituting \eqref{strange2} into \eqref{alp2} and then multiplying by $h^2(\bx)$, where $h(\bx)$ is an arbitrary real function of $\bx$ (but not $t$).  After space-time averaging we have:
\begin{equation}
s\la h^2 \ra   = \la h^2\gamma\ra  -\la h^2\bu\bcdot\bnabla \hat Z\ra - \kappa  \la  \bnabla h^2 \bcdot \bnabla \hat Z \ra + \kappa \la h^2{\left|\bnabla \hat Z \right|}^2\ra \, . \label{varCalc3}
\end{equation}
The first important consequence of insisting that $h(\bx)$ is independent of $t$ is that $\la h^2 \hat Z_t\ra=0$. Rearranging \eqref{varCalc3} yields
a quadratic in $h\bnabla{}Z$, and so we  complete the square:
\begin{equation}
s\la h^2 \ra  = \la h^2 \gamma \ra +  \kappa \la {\left|h\bnabla \hat Z - \bnabla h - \frac{h\bu}{2\kappa} \right|}^2 -   {\left|\bnabla h + \frac{h\bu}{2\kappa} \right|}^2 \ra \, . \label{varCalc6}
\end{equation}
Dropping the square term containing $\bnabla \hat Z$ from the right-hand side of \eqref{varCalc6} results in the inequality
\begin{equation}
s\la h^2 \ra  \geq \la  h^2\gamma\ra - \kappa \la \left| \bnabla h\right|^2  \ra - \quarter \kappa^{-1}  \la h^2 \left| \bu \right|^2 \ra \, . \label{varCalc7}
\end{equation}
Survival $(s>0)$ is guaranteed if we can find \textit{any} real function $h(\bx)$ which makes the right-hand side of \eqref{varCalc7} positive. 

To avoid guessing at $h(\bx)$ we apply variational calculus to \eqref{varCalc7} and find the function $\hs(\bx)$ which maximizes the right-hand side. Thus we apply the constraint
\begin{equation}
\la h^2 \ra=1
\label{Fconstraint}
\end{equation}
with a Lagrange multiplier $\tilde s$ 
and maximize the functional
\begin{equation}
\F[h] \equiv \la  h^2\gamma\ra - \kappa \la \left| \bnabla h\right|^2  \ra -\quarter \kappa^{-1}  \la h^2 \left| \bu \right|^2 \ra - \tilde{s} \la h^2 -1\ra \, .
\label{Fdef}
\end{equation}
The second important consequence of taking $h(\bx)$ independent of $t$ is that the time-average in $\la  h^2 |\bu|^2 \ra$ applies only to $|\bu|^2$,  so that for statistically homogeneous and isotropic flows (SHIF), such as the model in \eqref{renwave1}, $\la  h^2 |\bu|^2 \ra= \la |\bu|^2 \ra \la h^2 \ra = U^2 \la h^2 \ra$. Thus in this case
\begin{equation}
\F[h] \equiv \la  h^2\gamma\ra - \kappa \la \left| \bnabla h\right|^2  \ra -\quarter U^2 \kappa^{-1}  \la h^2  \ra - \tilde{s} \la h^2 -1\ra\, , \qquad \text{(for SHIF).}
\label{SHIF}
\end{equation}
The corresponding Euler-Lagrange equation is:
\begin{equation}
\frac{\delta \F}{\delta h}=0\, , \qquad\Rightarrow\qquad  \kappa \nabla^2 \hs + \left( \gamma - \frac{U^2}{4\kappa} \right) \hs = \tilde{s} \hs \, , \label{varCalc11}
\end{equation}
where $\hs(\bx)$ is the function which optimizes \eqref{varCalc7} by maximizing the right hand side.

\begin{figure*}
\begin{center}
\mbox{\includegraphics[width=.75\textwidth]{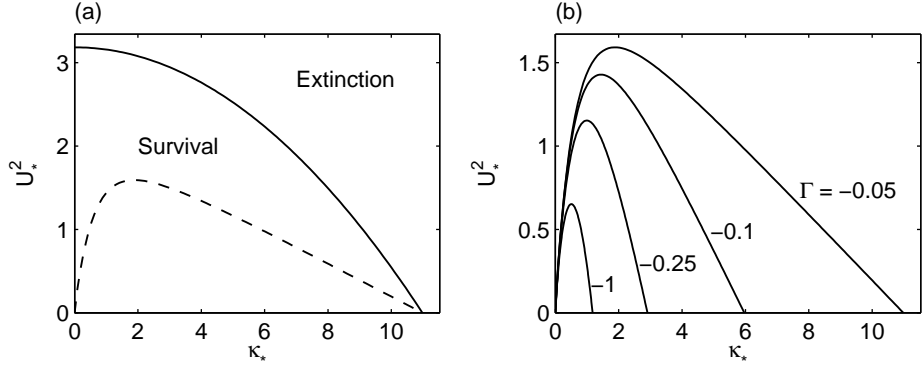}}
\end{center}
\caption{(a)  A schematic illustration; below the dashed curve the sufficient condition \eqref{varCalc12} guarantees survival.  However, the actual survival region may be much larger, as indicated  by the region below the  solid curve.  (b)  Below each solid curve the sufficient condition \eqref{varCalc12},  obtained by numerical solution of \eqref{varCalc11}, guarantees survival for the indicated value of $\Gamma$ in the sinusoidal growth model \eqref{sinugamma}.}
\label{extProhibFig}
\end{figure*}

Conveniently, if we multiply \eqref{varCalc11} by $\hs$ and average, we obtain
\begin{equation}
\tilde{s} =  \F\left[\hs\right] \, , \quad \text{and therefore } \quad s > \tilde{s}\, , \label{varCalc12}
\end{equation}
i.e., the population survives if the maximum eigenvalue $\tilde{s}$ of \eqref{varCalc11} is positive.  Note that \eqref{varCalc11} is the  same eigenproblem  solved in \eqref{zeroUeigen1} to determine the survival-extinction transition with $\bu=0$: the effect of $\bu \neq 0$ is the same as reducing the growth rate $\gamma(y)$ by $U^2/4 \kappa$.

The region below the dashed curve in  Figure \ref{extProhibFig}(a) is where \eqref{varCalc12} forbids extinction;  this region is contained within the actual survival region for our growth and stirring model, which is the area below the solid curve. Note that the survival region associated with the solid curve  in Figure \ref{extProhibFig}(a) abuts the $U_*^2$-axis, which is qualitatively different from the region below the dashed curve (see also Figure \ref{extProhibFig}(b)). The schematic solid curve in Figure \ref{extProhibFig}(a) is  based on our experience with simulations such as those in Figure \ref{strangeGrowthFig}  indicating that if $U_*$ is below some threshold, roughly $U_*=6$ in Figure \ref{strangeGrowthFig}(b), then the population survives in the limit $\kappa_* \to 0$ with $U_*$ fixed. This behavior is indicated in Figure \ref{strangeGrowthFig}(a): decreasing $\kappa_*$ with fixed $U_*=2$ increases $s$ towards Bayly's upper bound $\gamma_{\mathrm{max}}$.

The discrepancy between the actual survival region and the sufficient condition obtained from \eqref{varCalc7} depends on the details of the flow and growth rate.  For example, using the growth rate \eqref{sinugamma} and flow \eqref{renwave1}, the actual survival region is much larger than that calculated from \eqref{varCalc7}.  This is shown in Figure \ref{strangeGrowthFig}(a) where the plankton survive with $\Gamma = -1$, $U_*^2 = 4$ and $0.01 \leq\kappa_*\leq 1$; these parameter values are far above the curve $\Gamma  = -1$ in Figure \ref{extProhibFig}(b). Thus unfortunately \eqref{varCalc7} is not in general a tight lower bound on the survival exponent $s$.

\section{The statistical steady state \label{boundsSection}}

We now suppose that the population survives and turn to the statistical steady state which ensues once the quadratic nonlinearity in \eqref{sle} halts the exponential growth of the strange eigenmode. Two descriptors of this equilibrium are the biomass $\B$ and the productivity $\PP$ defined by
\begin{equation}
\B \equiv \la P \ra\, , \quad \text{and} \quad  \PP \equiv \la \gamma P \ra \, ,
\end{equation}
where $\la \ra$ is the space-time average defined in \eqref{eq1}. 
 An important integral constraint  is obtained by averaging \eqref{sle}:
\begin{equation}
 \la \gamma P \ra = \eta \la P^2 \ra \, .
 \label{eq3}
\end{equation}
Equation \eqref{eq3}  has the obvious interpretation that in statistical equilibrium reproduction is balanced by mortality.  Using \eqref{eq3} we see that the variance of $P(\bx,t)$, that is $\la P^2\ra - \la P\ra^2$, is equal to  $\eta^{-1} \PP - \B^2$. Thus $\B$ and $\PP$ provide the mean and variance of $P(\bx,t)$.  This is a strong motivation for regarding $\B$ and $\PP$ as the most fundamental statistical descriptors of the system and for attempting to understand their dependence on  $\kappa$ and the properties of $\bu(\bx, t)$ and  $\gamma(\bx, t)$.

We obtain a  second integral constraint by making the change of variables $Z \equiv \ln P$ in \eqref{sle}:
\begin{equation}
Z_t + \bu\bcdot\bnabla Z = \gamma(\bx,t) - \eta P + \kappa \left( \nabla^2Z + {\left|\bnabla Z \right|}^2 \right) \, . \label{Zeqn}
\end{equation}
Averaging \eqref{Zeqn} gives the equilibrium analog of \eqref{alp4}:
\begin{equation} 
\eta \la  P \ra = \la \gamma \ra + \kappa \la |\bnabla Z|^2 \ra\, .
\label{lb1}
\end{equation}
This shows that  the $\B$ is always greater than $\eta^{-1} \la \gamma \ra$. This lower bound is only useful if $\la \gamma \ra >0$, so 
\begin{equation}
 \eta \B \geq \max\left\{0, \la \gamma \ra  \right\}    \, .
\label{lb1.1}
\end{equation}

To obtain a lower bound on the productivity we combine the Cauchy-Schwarz inequality,  $\la{}P^2 \ra \geq \la P \ra^2$, with the definition of $\PP$ and the identity \eqref{eq3} to obtain
$
 \la \gamma \ra \times \max\left\{0, \la \gamma \ra \right\} \leq \eta \PP 
$.
We can also employ Cauchy-Schwarz  to find an upper bound on the productivity:  $\PP \equiv \la \gamma P \ra \leq \sqrt{\la \gamma^2\ra \la P^2\ra}$. Using \eqref{eq3} to replace $\la P^2 \ra$ by $\eta^{-1} \PP$, and squaring the resulting inequality, we obtain the upper bound  $\eta \PP \leq  \la \gamma^2 \ra$. Thus, to summarize:
\begin{equation}
\la \gamma \ra \times \max\left\{0, \la \gamma \ra \right\} \leq \eta \PP \leq  \la \gamma^2 \ra\, .
 \label{lb1.2}
 \end{equation}

 The simple bounds in \eqref{lb1.1} and \eqref{lb1.2} involve neither $\bu(\bx,t)$ nor $\kappa$. In the next section we obtain a more elaborate bound which depends on $\bu(\bx,t)$ and $\kappa$ and which applies to the case $\la \gamma \ra <0$.

\subsection{A second lower bound on the biomass\label{lowerSection2}}

\begin{figure*}
\begin{center}
\mbox{\includegraphics[width=.75\textwidth]{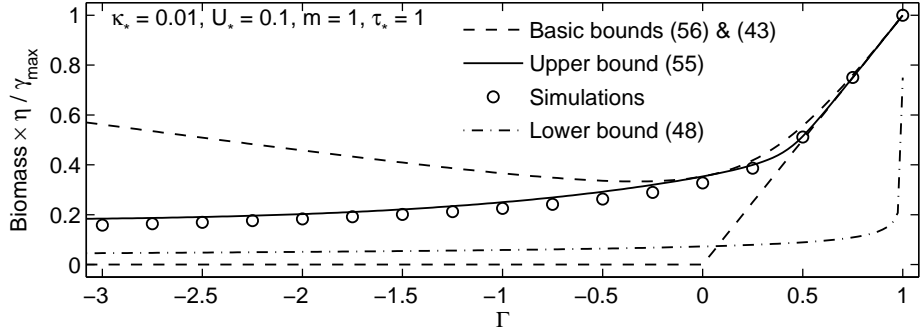}}
\end{center}
\caption{The dashed curves are the basic lower \eqref{lb1.1} and upper \eqref{simpleUp1}  bounds: $\max\left\{0, \la \gamma \ra  \right\} \leq \eta \B \leq  (\sqrt{\la \gamma^2 \ra} + \la \gamma \ra)/2$.  The solid curve is the optimized upper bound \eqref{up8} and the dash-dot curve is the lower bound \eqref{varLower10}.  The circles are the biomass obtained from simulations.  The  lower bound \eqref{varLower10} is tighter than the basic bound \eqref{lb1.1} only for $\Gamma\lesssim 0.075$.}
\label{BioBoundsFig}
\bigskip
\end{figure*}

To obtain a lower bound on $\B \equiv \la P \ra$ we follow the calculation in subsection \textit{\ref{prohibition}}. Multiplying \eqref{Zeqn} by $h^2(\bx)$ and space-time averaging we obtain a steady state analog of \eqref{varCalc3}:
\begin{equation}
\eta \la h^2 P \ra   = \la h^2\gamma\ra  -\la h^2\bu\bcdot\bnabla  Z\ra - \kappa  \la  \bnabla h^2 \bcdot \bnabla  Z \ra + \kappa \la h^2{\left|\bnabla  Z \right|}^2\ra \, . \label{varCalc3'}
\end{equation}
Repeating the manipulations in \eqref{varCalc6} and \eqref{varCalc7}, and again assuming that $\bu(\bx,t)$ is statistically homogeneous and isotropic, we obtain
\begin{equation}
\eta \frac{\la h^2 P \ra}{\la h^2 \ra} \geq \F[h]\, ,
\end{equation}
where $\F(h)$ is the functional defined in \eqref{SHIF}. Thus maximizing $\F(h)$ by solving the Euler-Lagrange equation \eqref{varCalc11} for $\hs$ does double duty: we obtain both a lower bound on the survival exponent $s$ and on the ratio $\eta {\la h^2 P \ra}/{\la h^2 \ra}$. Noting that
\begin{equation}
\max(h^2) \la P \ra \geq \la h^2 P \ra\, , \label{varCalc5}
\end{equation}
and using the normalization $\la h^2 \ra=1$, this lower bound also provides a lower bound on the  biomass $\B$:
\begin{equation}
\eta \B \geq  \frac{\F\left[\hs\right]}{\max\left(\hs^2\right)} \, . \label{varLower10}
\end{equation}
Unfortunately the inequality in \eqref{varCalc5} is crude, and so the lower bound \eqref{varLower10} is not always tighter than the basic lower bound in \eqref{lb1.1}.  This is shown in Figure \ref{BioBoundsFig} where \eqref{varLower10} is an improvement over \eqref{lb1.1} only for $\Gamma \lesssim 0.075$.

\subsection{An upper bound on the biomass\label{upperSection}}

Having found lower bounds on the biomass in \eqref{lb1.1} and \eqref{varLower10}
 we now seek a complementary upper bound.  The first step is to obtain a constraint by multiplying \eqref{sle} by an arbitrary positive function $f(\bx)$ and averaging:
\begin{equation}
\la \eta{}fP^2 - gP\ra = 0 \, , \label{up1}
\end{equation}
where
\begin{equation}
g(\bx,t) \equiv \bu\bcdot\bnabla{}f + \kappa\nabla^2{}f + f\gamma \, . \label{up2}
\end{equation}
Notice that $g(\bx,t)$ inherits time dependence from $\bu(\bx,t)$ and $\gamma(\bx,t)$.

We add $\beta$ times \eqref{up1} onto the definition of $\B$
\begin{equation}
\B = \la P \ra + \beta\la \eta{}fP^2 - gP\ra \, , 
\end{equation}
and then complete the square:
\begin{align}
\B &=\la \frac{{(\beta g + 1)}^2}{4\beta\eta{}f} \ra - \beta \la \eta{}f{\left(P -\frac{\beta{}g+1}{2\beta\eta{}f} \right)}^2\ra \, , \label{fstline} \\
&\leq  \la \frac{{(\beta g + 1)}^2}{4\beta\eta{}f} \ra\, .
\label{completesquare1}
\end{align}
In passing from the first to the second line we assume that $\beta>0$ so that dropping the final term in \eqref{fstline} results in an upper bound on $\B$.
Minimizing the right-hand side of the inequality \eqref{completesquare1} yields the tightest bound.  The optimal value of $\beta$ is 
\begin{equation}
\beta_* = \sqrt{\la \frac{1}{f}\ra {\la \frac{g^2}{{}f}\ra}} \, ,
\end{equation}
which is non-negative and therefore consistent with \eqref{completesquare1} being an upper bound.  Substituting $\beta_*$ into \eqref{completesquare1} yields
\begin{equation}
\B \leq \frac{1}{2\eta}\left[ \la \frac{g}{f}\ra + \sqrt{\la \frac{1}{f}\ra {\la \frac{g^2}{f}\ra}}\, \right] \, . \label{up8}
\end{equation}

For the simple choice $f(\bx) = 1$, which implies $g(\bx,t)=\gamma(\bx,t)$, \eqref{up8} immediately delivers the upper bound
\begin{equation}
\B \leq \frac{\la \gamma \ra + \sqrt{ \la \gamma^2 \ra}}{2\eta}  \, . \label{simpleUp1}
\end{equation}
The upper bound above is sharp in the special case of constant $\gamma$ where the stable attracting state is $\B = \gamma/\eta$ and $\PP=\gamma^2/\eta$.

To improve on \eqref{simpleUp1} requires a better comparison function than $f(\bx) =1$. One can attempt to optimize the choice of $f(\bx)$ by maximizing the the right hand side of \eqref{up8} using variational calculus.  Unfortunately the resulting Euler-Lagrange equation is very complicated and so we compromise by using a simple trial function, such as 
\begin{equation}
f = \ee^{ p\cos k_1 y }\, .\label{trialf}
\end{equation}
The adjustable parameter $p$ is determined by minimizing the right-hand side of \eqref{up8}. Since $p=0$ corresponds to $f=1$ this procedure can only improve on \eqref{simpleUp1}. This trial function procedure has been implemented numerically using the sinusoidal $\gamma(y)$ in \eqref{sinugamma} to obtain the upper bound indicated by the solid line in Figure \ref{BioBoundsFig}.   In Figure \ref{BioBoundsFig} this bound is very tight and is the bound closest to the simulation results.

\subsection{Summary of the bounds}

\begin{figure*}
\begin{center}
\mbox{\includegraphics[width=.75\textwidth]{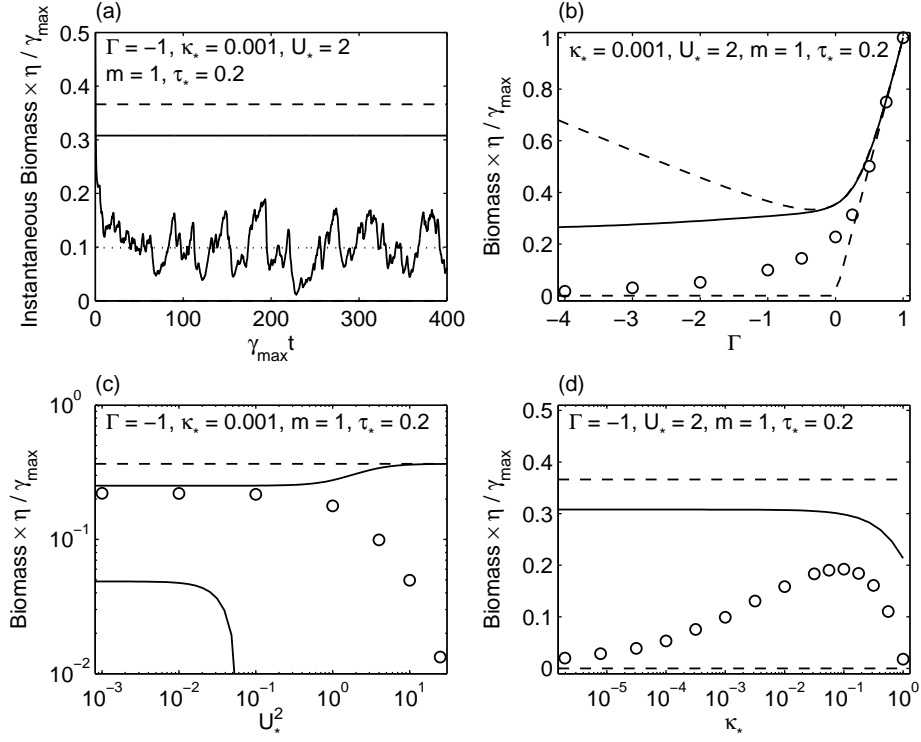}}
\end{center}
\caption{(a)  A time series of $\bar P(t)$ compared with two upper bounds;  the dashed line is the simple upper bound in \eqref{summary} and the straight solid line is the optimized upper bound in \eqref{summ2} using the trial function $f = \exp{[0.3960\cos(k_1 y)]}$.  The dotted line shows the average of the biomass from $t_*=100$ to $400$.  (b)-(c)  The open circles indicate the biomass obtained from numerical simulation for the indicated parameter values.  The dashed lines are the upper and lower bounds in \eqref{summary} and the solid lines are the bounds in \eqref{summ2} with the upper bound $p$-optimized as in \eqref{trialf}.   The lower bound in \eqref{summ2} only appears in (c) where it is not very tight but does forbid extinction for small $\Ustar$.}
\label{basicBoundsFig}
\bigskip
\end{figure*}

Our simplest bounds on the biomass are
\begin{equation}
\max\left\{0, \la \gamma \ra  \right\} \leq \eta \B \leq \frac{1}{2} \left[ \sqrt{\la \gamma^2 \ra} + \la \gamma \ra\right]  \, .
\label{summary}
\end{equation}
We also derived two bounds involving $\kappastar$, $\Ustar$ and the details of $\gamma$:
\begin{equation}
\frac{\F\left[\hs\right]}{\eta\max\left(\hs^2\right)}   \leq B \leq \frac{1}{2\eta}\left[ \la \frac{g}{f}\ra + \sqrt{\la \frac{1}{f}\ra {\la \frac{g^2}{f}\ra}}\right] \, , \label{summ2}
\end{equation}
where $\hs$ is the solution to \eqref{varCalc11} and $g$ is defined in \eqref{up2}.  The bounds in \eqref{summ2} can be tighter than the bounds in \eqref{summary}, but the bounds in \eqref{summary} have the advantage that they only require knowing the mean and variance of $\gamma$, while the more complicated bounds require knowing $\gamma$ everywhere.

The productivity is subject to simpler constraints:
\begin{equation}
\la \gamma \ra \times \max\left\{0, \la \gamma \ra \right\} \leq    \eta \PP \leq \la \gamma^2 \ra \, .
\label{prodSummary}
\end{equation}

We evaluate and illustrate the bounds \eqref{summary} and \eqref{summ2} in Figure \ref{basicBoundsFig}.  Panel (a) shows an example of a time series of $\bar{P}(t)$ in an adverse average environment  ($\Gamma =-1$);  the optimized upper bound in \eqref{summ2}, indicated by the solid curve, is obtained by optimizing the trial function in \eqref{trialf}  over $p$.  With $\Ustar=2$, a largish value, even the optimized upper bound \eqref{summ2} is generously large and the lower bound in \eqref{summ2} is useless --- the best lower bound in this case is simply $0 \leq \B$. 

Figure \ref{basicBoundsFig}(b) shows that both of the bounds in \eqref{summary} and the optimized upper bound in \eqref{summ2} become tight as $\Gamma \to 1$ and the growth rate becomes spatially uniform. However, the growth rate does not need to be very uniform for the bounds to work well.  For example, with $\Gamma = 0.5$ the growth rate $\gamma(y)$ varies from $0$ to $\gammam$, but despite this large spatial inhomogeneity the lower bound in \eqref{summary} and the upper bound in \eqref{summ2} constrain the biomass to lie between $0.5\gammam/\eta$ and $0.5508\gammam/\eta$.

The effects of stirring on the biomass  are shown in \ref{basicBoundsFig}(c): increasing $\Ustar$ loosens the bounds and sufficiently large $U_*$ drives the plankton to extinction.  The optimized upper bound in \eqref{summ2} is tight provided that $U_*$ is not too large and the lower bound in \eqref{summ2} only appears for very small $\Ustar$.   

Figure \ref{basicBoundsFig}(d) shows that there is a value of $\kappastar$ (roughly  $\kappa_*\approx 10^{-1}$) at which the biomass is maximized. Unfortunately  the optimized upper bound in \eqref{summ2} shows dependence on $\kappastar$ only when diffusion is rather strong and the lower bound in \eqref{summ2} does not appear and so neither of the bounds containing $\kappastar$ are delicate enough to indicate the existence of a maximum biomass at a particular value of $\kappa_*$.

Finally, the overall the performance of the bounds in Figure \ref{basicBoundsFig} is worse than in Figure \ref{BioBoundsFig}  because the simulations in Figure \ref{basicBoundsFig} use much larger values of $\Ustar$. The main conclusion is that our bounds on $\B$ are tight if $U_* \lesssim O(1)$ and  $\Delta \gamma/\gamma \lesssim O(1)$. In the next section we use \eqref{eq3} and \eqref{lb1} to obtain joint inequalities  constraining $\B$ and $\PP$.

\section{Simultaneous Bounds on $\B$ and $\PP$ \label{jointSection}}

The only information concerning $\PP$ provided by the bounds in the previous section is \eqref{prodSummary}.
In this section we supplement \eqref{summary} and \eqref{prodSummary} by deriving bounds  involving $\B$ and $\PP$ together.  The first simultaneous bound is obtained by observing that $\la \gamma P \ra \leq \gammam \la P \ra$, where $\gammam$ is the global maximum of $\gamma (\bx,t)$.  Therefore
\begin{equation}
\PP \leq \gammam\B \, . \label{simul1}
\end{equation}
We obtain a more elaborate simultaneous bound by adding two versions of zero to the definition of $\PP$:
\begin{eqnarray}
\PP &= &\la \gamma P\ra+ \alpha \left[\B - \la P \ra \right] + \beta \left [\la \gamma P \ra - \eta \la P^2 \ra \right]\, , \nonumber \\
 &= & \alpha \B- \beta \eta \la\left[P - \frac{\gamma + \beta \gamma - \alpha}{2 \beta \eta} \right]^2 \ra + \frac{ \la(\gamma + \beta \gamma - \alpha)^2 \ra}{4 \beta \eta}  \, . 
 \label{BP1}
 \end{eqnarray}
 Above $\alpha$ and $\beta$ are constants.

 If $\beta >0$, then  we obtain an upper bound on the productivity by dropping the squared  term containing $P$ from the right-hand side \eqref{BP1}. Optimizing this upper bound by minimizing over both $\alpha$ and $\beta$ (Appendix \ref{optApp}) gives
 \begin{equation}
 \PP \leq \la \gamma \ra \B + \frac{\sigma^2}{2 \eta} + \sigma \sqrt{\frac{\la \gamma^2 \ra}{4 \eta^2} -\left(\B  - \frac{\la \gamma\ra}{2 \eta}\right)^2}\, ,
 \label{BPupper}
 \end{equation}
 where 
 $$
 \sigma^2 \equiv \la \gamma^2\ra - \la \gamma \ra^2\, .
 $$

\begin{figure*}
\begin{center}
\mbox{\includegraphics[width=.75\textwidth]{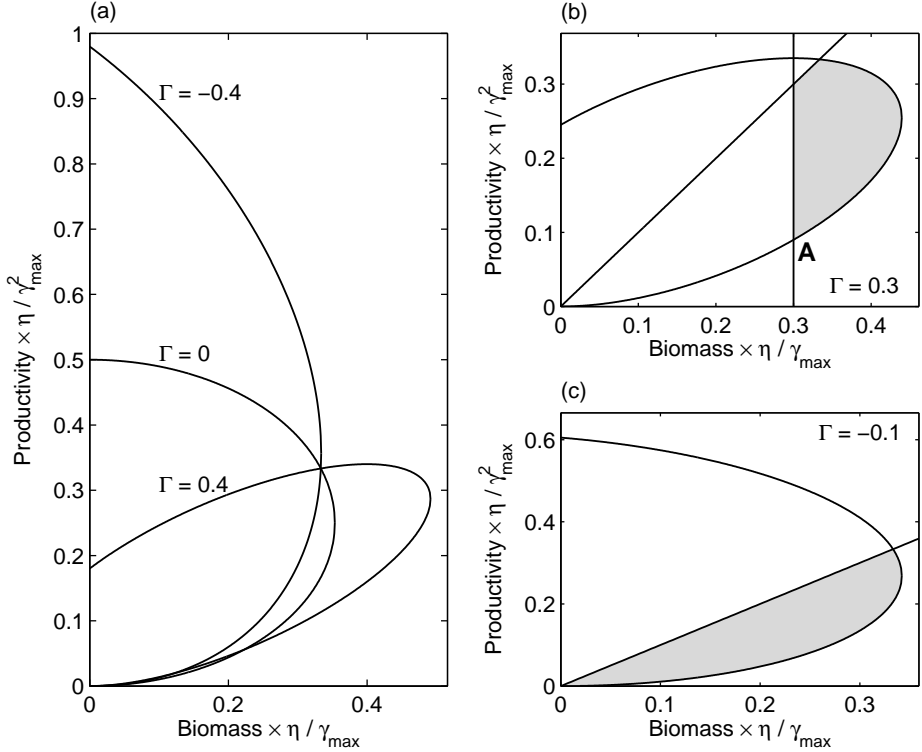}}
\end{center}
\caption{(a)  Bounding ellipses obtained from \eqref{double} and  the sinusoidal model in \eqref{sinugamma}; units are non-dimensional with $\gammam=\eta =1$.  As $\Gamma$ decreases the growth rate becomes more spatially inhomogeneous  and the ellipse expands and the simultaneous bounds become less tight.  However, because $\PP \leq \gamma_{\mathrm{max}} \B$ much of the expanded ellipse is inaccessible.  (b)  The shaded region indicates the allowed region if $\la \gamma\ra >0$. The   point $(\B,\PP) = \eta^{-1} (\la \gamma\ra, \la \gamma \ra^2 )$ is indicated by $A$.  (c)  The shaded region indicates the allowed region if $\la \gamma\ra < 0$. In this case the point $A$ lies outside the first quadrant and so the constraint \eqref{prodSummary1} is toothless. }
\label{ellipzfig1}
\end{figure*}

Returning to (\ref{BP1}) and  taking $\beta < 0$  we obtain a complementary lower bound by again dropping the squared term containing $P$. Maximizing over $\alpha$ and $\beta$ gives an expression analogous to \eqref{BPupper}, except that the inequality  and the sign of the square root are reversed. Thus we have proved that 
 \begin{equation}
 \PP_- (\B)  \leq  \PP \leq \PP_+(\B)   \label{double}\, ,
\end{equation}
where the functions $ \PP_- (\B)$ and $ \PP_+ (\B)$ are
\begin{equation}
\PP_{\pm} (\B) \equiv \la \gamma \ra \B + \frac{\sigma^2}{2 \eta} \pm \sigma \sqrt{\frac{\la \gamma^2 \ra}{4 \eta^2} -\left(\B  - \frac{\la \gamma\ra}{2 \eta}\right)^2}\, .
\label{doubledef}
\end{equation}
 The inequality above constrains the system to fall within the intersection of the first quadrant of the $(\B,\PP)$-plane and the ellipse defined by the arcs $\PP_+(\B)$ and $\PP_-(\B)$  --- see Figure \ref{ellipzfig1}(a).  
 
  Introducing the quantities
 \begin{equation}
 \B_+ \equiv \frac{\sqrt{\la \gamma^2 \ra} + \la \gamma \ra}{2 \eta}\, , \qquad 
 \B_- \equiv \frac{\sqrt{\la \gamma^2 \ra} - \la \gamma \ra}{2 \eta}\, ,
 \label{doubledef1}
 \end{equation}
 and noting from \eqref{simpleUp1} that $\B \leq \B_+$, 
we rewrite \eqref{doubledef} as
\begin{equation}
\PP_{\pm} (\B) = \eta  \left[\sqrt{\B_+(\B_- + \B)} \pm \sqrt{\B_- (\B_+ - \B)} \right]^2\, .
\label{doubledef2}
\end{equation}
Taking $\PP_-(\B)$, the right hand side of \eqref{doubledef2} has a double root at $\B=0$. Thus the $\B$-axis is tangent to the  arc $\PP= \PP_-(\B)$ at the origin of the $[\B,\PP]$-plane (e.g, see Figure \ref{ellipzfig1}).

In \eqref{summary} and \eqref{prodSummary} we obtained the lower bounds
\begin{equation}
\max\left\{0, \la \gamma \ra \right\} \leq \eta \B \quad \text{and} \quad 
\la \gamma \ra \times \max\left\{0, \la \gamma \ra  \right\} \leq    \eta \PP\, .
\label{prodSummary1}
\end{equation}
Employing \eqref{prodSummary1}, and the upper bound $\PP \leq \gammam \B$, in  concert with \eqref{double}\footnote{The lower bound $\eta \B^2 \leq \PP$ obtained from \eqref{eq3} and Cauchy-Schwarz is weaker than \eqref{double}: the parabola $\PP = \eta \B^2$ is below the elliptic arc $\PP = \PP_-(\B)$ everywhere, except at the point  $\eta (\B,\PP) = (\la \gamma\ra/\eta, \la \gamma \ra^2) $, which is $A$ in Figure \ref{ellipzfig1}. At $A$ the two curves touch and share a tangent.} we restrict the system to the shaded region in Figures \ref{ellipzfig1}(b) and (c). Specifically, the system must fall within the  intersection of:
 \begin{enumerate}
 \item the first quadrant of the $(\B,\PP)$-plane;
 \item  the interior of the bounding ellipse;
 \item  the wedge  $0< \PP \leq \gammam \B$;
  \item   the lower bounds in \eqref{prodSummary1}, which define a quadrant with   southwest apex $A$ in Figure \ref{ellipzfig1}(b).
 \end{enumerate}

 These joint constraints --- which use only $\la \gamma \ra, \la \gamma^2 \ra, \gammam$ and $\eta$ ---  provide  basic information about the possible range of the biomass and productivity.  
 
  \begin{figure*}
\begin{center}
\mbox{\includegraphics[width=.75\textwidth]{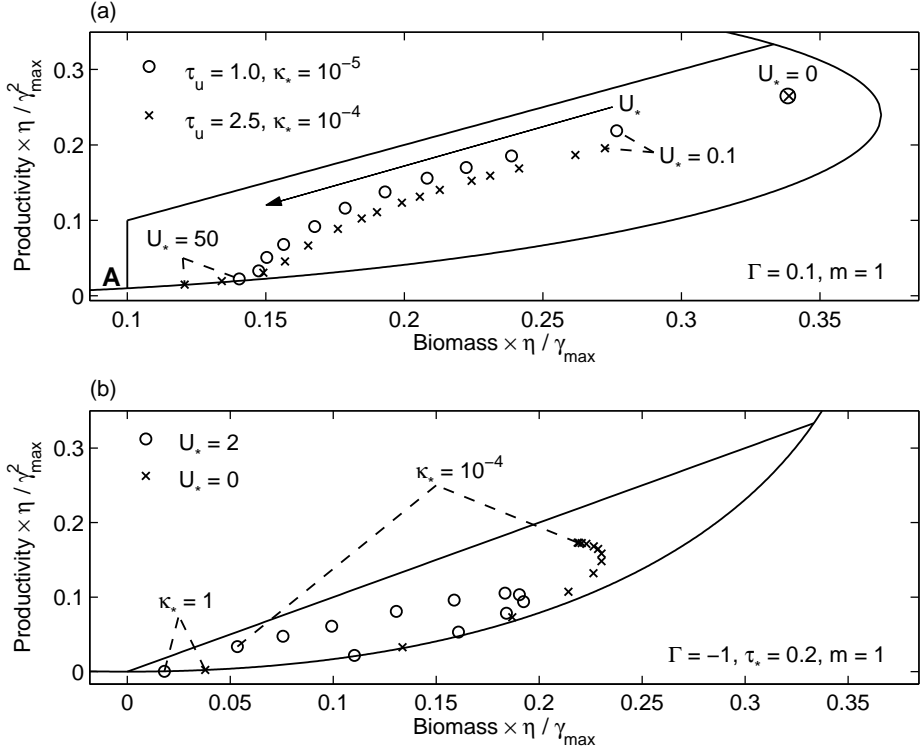}}
\end{center}
\caption{(a)  The joint bounds illustrated with variable $\Ustar$ and fixed $\tau_u$.  $U_*$ increases from $0.1$ to $50$ (b)  The joint bound illustrated with variable $\kappastar$; the maximum value of $\kappastar$ is $1$ for  both $U_*=0$ and $2$.}
\label{BPBoundsFig}
\end{figure*}

 Figure \ref{BPBoundsFig} illustrates the joint bound.  In panel (a) the parameter $\Ustar$ is varied by a factor of $500$ while $\tau_u \equiv U k_m \tau$ in \eqref{LyapDef1} is  fixed. Increasing $\Ustar$ decreases both the biomass and the productivity so that for large $\Ustar$ the system is very near the lower arc of the bounding ellipse $\PP_-$ in \eqref{doubledef}.  Panel (b) shows the effect of varying $\kappastar$.  $\Gamma < 0$ in this case and so large $\kappastar$ drives the system to the origin (extinction).  Small $\kappastar$ also causes the system to head towards extinction when $\Ustar = 2$.  For $\Ustar = 0$ the system heads to the local solution $P(y) = \gamma(y)/\eta$ as $\kappastar\to 0$.  For both values of $\Ustar$ the maximum biomass is achieved with a moderate value of $\kappastar$:  about $0.04$ and $0.1$ for $\Ustar = 0$ and $2$, respectively.  The maximum values of the biomass corresponding to these parameter values are about $0.23$ and $0.19$.

\section{Conclusions and discussion \label{conclusions}}

Previous work on the FKPP equation and its relatives focused on the survival-extinction transition \cite{KiersteadS1953, DahmenNS2000, LinMT-OLKS2004, BJBayly1992}, filamentation \cite{NeufeldLH1999, NeufeldHT2002}, and front propagation \cite{Fisher1937, KPP1937, AudolyBP2000}.  Here we consider the survival-extinction transition and find the existence of a survival region in diffusivity-velocity parameter space even if the average growth rate is negative.  After considering the survival-extinction transition we move on to estimating the average and variance of the concentration of plankton (chemical products) in the domain by deriving upper and lower bounds on the biomass and productivity.

Our bounds on the biomass and productivity apply to any plankton model (or chemical reaction) with a linear growth term, quadratic damping, and periodic or no-flux boundary conditions.  The simplest bounds in \eqref{summary} do not require detailed knowledge of the growth rate, only the mean and variance.  Furthermore, the bounds in \eqref{summary} do not use information about diffusion or the velocity field.  This may be attractive, depending on whether or not the information is available.

If the flow is statistically homogeneous and isotropic and $\la |\bu|^2\ra$, the diffusivity and the details of the plankton growth rate are known, then we can find more elaborate and possibly tighter bounds on the biomass.  The upper bound obtained via the trial function procedure \eqref{summ2} is always tighter than the upper bound in \eqref{summary} while the lower bound in \eqref{varLower10} may be tighter than that in \eqref{summary}.  Additionally, even when the bound \eqref{varLower10} is not tight, it may forbid extinction, as in Figures \ref{BioBoundsFig} and \ref{basicBoundsFig}(c).

In addition to constraining the biomass, we also derived bounds on the productivity.  Our definition of productivity, $\PP \equiv \la\gamma{}P\ra$, is unusual as it includes regions where the growth rate $\gamma$ is negative.  Regardless, it is still useful to bound this quantity because from \eqref{eq3}, $\la\gamma{}P\ra$ equals the mean-squared plankton concentration and may be combined with the biomass to find the variance of the plankton concentration.

Finally, we derived simultaneous bounds on the biomass and productivity which constrain the system to lie inside a certain portion of the biomass-productivity plane.  The system is close to the boundary of this region for moderate to large diffusivity and velocity.  These bounds may be useful as an alternative to parameterizing sub-grid scale processes in ecological model and for predicting the results of chemical reactions.

\begin{figure*}
\begin{center}
\mbox{\includegraphics[width=.75\textwidth]{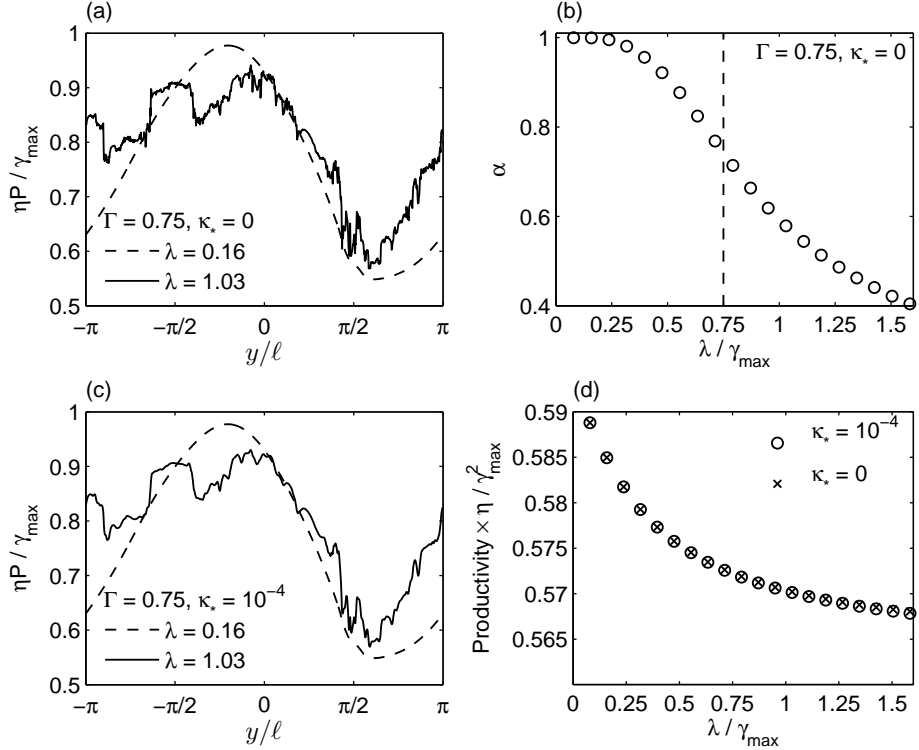}}
\end{center}
\caption{(a)  Transects along the line $x=0$ for two solutions of \eqref{sle} with $\kappastar = 0$ using the same sequence of phases $\phi_{x}$ and $\phi_y$ in \eqref{renwave1} but different values of $\lambda$ in \eqref{LyapDef}.  $\lambda$ is varied by holding $\tau_u$ and $m$ fixed and varying $\Ustar$.  The transect with $\lambda = 0.16$ is smooth, but the transect with $\lambda = 1.03$ is not.  (b)  The H\"{o}lder exponent \eqref{holderDef}.  The vertical dashed line is the line $\lambda = \Gamma$ and coincides with the inflection point of $\alpha(\lambda)$.  (c)  Transects along the line $x=0$ for two solutions of \eqref{sle} with $\kappastar = 10^{-4}$.  The $\lambda = 0.16$ transect is very similar to the corresponding transect in (a), but the $\lambda = 1.03$ transect is noticeably smoother.  (d)  The productivity $\PP$ as a function of $\lambda$.}
\label{filamentsFig}
\end{figure*} 

In concluding, we note that the filamentation transition discussed by Neufeld, L\'{o}pez and Haynes \cite{NeufeldLH1999} appears in our model if we take $\Gamma > 0$ and make the simplification $\kappa = 0$.  Examining this transition allows us to assess the effect of small-scale structure on the productivity (the biomas is constrained to equal $\Gamma$ by \eqref{lb1}).The filamentation transition occurs when the flow Lyapunov exponent, $\lambda$ in \eqref{LyapDef}, exceeds the rate of damping back to local equilibrium and $P$ develops narrow filaments which are not smooth in the direction orthogonal to the filaments.  If the damping back to local equilibrium is strong enough then no filaments form and the field remains smooth even if $\kappa=0$.  Figure \ref{filamentsFig}(a) shows transects along the line $x = 0$ for two different runs using the same sequence of  phases $\phi_x$ and $\phi_y$ in \eqref{renwave1} and  the same parameter values, except for $\Ustar$ and $\tau_*$.  The argument of the Lyapunov function $\tau_u$ in \eqref{LyapDef1} is held fixed and so the Lyapunov exponent of the flow increases linearly with $\Ustar$.  For the case $\lambda = 0.16$ the plankton distribution is smooth, but for $\lambda = 1.03$ the distribution is filamented with many sharp peaks in the transect, indicating that we have passed throught the filamentation transition.  

Figure \ref{filamentsFig}(b) shows the H\"older exponent $\alpha$, which is defined in \cite{NeufeldLH1999} as
\begin{equation}
{\delta P}\sim{|\delta\bx|}^\alpha \, . \label{holderDef}
\end{equation}
In our model $\alpha\approx 1$ for $\lambda < 0.25$ and then smoothly decreases to about $0.4$ for $\lambda \approx 1.6$.  In between $\lambda = 0.25$ and $\lambda\approx 1.6$ the system transitions from smooth to filamented with the inflection point of $\alpha(\lambda)$ occurring at $\lambda = \Gamma$, as shown by the dashed line in Figure \ref{filamentsFig}(b).

Figure \ref{filamentsFig}(c) is the same as panel (a) except $\kappastar$ is small but non-zero.  The effect on the previously smooth case ($\lambda = 0.16$) is imperceptible, but the previously filamented case ($\lambda = 1.03$) is noticeably smoother.  Panel (d) shows that the productivity $\PP$ decreases as $\lambda$ increases .  The biomass $\B$ is always $0.75$ for $\kappastar = 0$ by \eqref{lb1} and constrained by \eqref{summary} to lie between $0.75$ and $0.7603$ for $\kappastar = 10^{-4}$.  For the values of $\lambda$ used in Figure \ref{filamentsFig} $\B$ varies by less than $1\%$.  Therefore, the variance, given by $\eta^{-1} \PP - \B^2$, decreases as the Lyapunov exponent increases and the field becomes less smooth.  This counterintuitive result occurs because with increased stirring the plankton feel the average growth rate and carrying capacity and consequently the magnitudes of the deviations from the mean are decreased as shown in \ref{filamentsFig}(a) and (c).  The filamentation transition has no effect on the biomass or the productivity in our model and this may be seen by comparing panels (b) and (d).  In panel (b) the H\"older exponent remains approximately $1$ until $\lambda \approx 0.25$, while the productivity in panel (d) begins dropping immediately.  This insensitivity to small scale structure indicates why crude inequalities, like the ones developed here, may be able to successfully constrain gross statistics such as $\B$ and $\PP$.

\section*{Acknowledgements}

This work was supported by the National Science Foundation grant  OCE-0220362.

\appendix

\section{The Lyapunov exponent \label{LyapExpAppen}}

Under the action of the velocity field \eqref{renwave1}, a particle starting at $(x,y)$ at $t=0$ moves to the point $(x',y')$ at $t=\tau$. This area-preserving transformation can be written down explicitly, and one can then calculate the Jacobian matrix which has the form:
\begin{equation}
J_1 \equiv \begin{pmatrix} x'_x & x'_y \\
y'_x & y'_y \end{pmatrix} = \begin{pmatrix} 1& \beta s_1 \\
\beta s_2 & 1+ \beta^2 s_1s_2 \end{pmatrix}\, , 
\end{equation} 
where $\beta \equiv U k_m \tau/\sqrt{2}$ and $s_n \equiv \sin \theta_n $, with $\theta_n$  a random phase. We denote the transpose of $J$ by $J^{\intercal}$.

Consider an ensemble of infinitesimal line elements; at $t=0$ each element  has  length $\ell_0$, and is oriented along a randomly directed unit vector $\be$. After one iteration, an element is stretched by a random factor $ \be^ {\intercal}  J_1^ {\intercal}  J_1 \be$ to a length $\ell_1 = \sqrt{\be^ {\intercal}  J_1^ {\intercal}  J_1 \be}\,  \ell_0$. After $n$ iterations the length is given by 
\begin{equation}
\ell_n = \sqrt{\be^ {\intercal}  J_n^{\intercal} \cdots J_1^ {\intercal}  J_1 \cdots J_n \be}\,  \ell_0 \, .
\end{equation}
The matrix $\J_n \equiv \left(J_1\cdots J_n\right)^{\intercal}  \left(J_1\cdots J_n\right)$ is real and symmetric with determinant one. Hence we have the representation
\begin{equation}
\J_n = Q^{\intercal}  E Q\, ,
\end{equation}
where the matrix $Q$ corresponds to  rotation through a random angle $\chi$ and  
\begin{equation}
E = \begin{pmatrix} \alpha & 0 \\ 0 & \alpha^{-1} \end{pmatrix}\, .
\end{equation}
$\alpha>1$ is an eigenvalue of $\J_n$.
The Lyapunov exponent $\lambda$ defined in \eqref{LyapDef} is given by $\lambda = \la \ln(\ell_n/\ell_0) \ra/(n\tau)$, where $\la \ra $ indicates an average over the random angles in $\be$ and $Q$ and over the the random eigenvalue $\alpha$. We cannot calculate this average analytically because  the average over $\alpha$ is complicated. However the average over the angles is  a standard integral:
\begin{align}
\la \ln(\ell_n/\ell_0) \ra_{\chi} &= \half \la  \ln \left(\alpha \cos^2 \chi +\alpha^{-1} \sin^2 \chi  \right)\ra_{\chi}\, , \\
&= \ln\left[\half \left(\alpha^{1/2} + \alpha^{-1/2}\right) \right]\, .
\label{monte}
\end{align}
An efficient Monte Carlo procedure takes advantage of \eqref{monte} by generating many realizations of $\J_n$, calculating $\alpha$ for each realization and obtaining the Lyapunov exponent as
\begin{equation}
\lambda = \frac{1}{n \tau} \la \ln\left[\half \left(\alpha^{1/2} + \alpha^{-1/2}\right) \right]
 \ra\, .
\end{equation} 
This procedure using $4\times10^4$ realizations of $\J_n$ with $n = 20$ gives the solid curve in Figure \ref{LyapFig}(a).

\section{Approximate solutions of a Mathieu eigenproblem \label{extinctApp}}

In section \textbf{3} we solve a variety of eigenproblems related to Mathieu's equation, written here on the domain $-\pi<y<\pi$ as
\begin{equation}
KQ_{yy} + ( \cos y- E)Q = 0 \, .
\label{mathieu2}
\end{equation}
This is not the standard form of Mathieu's equation, but it is  convenient for the problems we face in section \textbf{3}. Requiring that the solution of \eqref{mathieu2} be periodic determines an eigenrelation $K(E)$ between the parameters $K$ and $E$. In this appendix we focus on the first mode (that is the ground state) and obtain the main features of the function $K(E)$  via perturbation theory pivoted round the complementary limits $K \to 0$ and $K\to \infty$. This result is used to deduce the simple analytic approximations indicated by the dashed curves in Figure \textbf{\ref{zeroUFig}}(b).

If $K \gg 1$ we define $\epsilon \equiv K^{-1} \ll 1$ and regard $E(\epsilon)$ as an eigenvalue.  We then solve \eqref{mathieu2} via a regular perturbation expansion in $\epsilon$. The result is
\begin{equation}
Q =  1 + \epsilon \cos y +\frac{ \epsilon^2}{8} \cos 2 y + O(\epsilon^3)\, , \qquad E = \frac{\epsilon}{2} + O(\epsilon^3)\, ,
\end{equation}
or 
\begin{equation}
K \approx \frac{1}{2E}\, , \quad \text{if $K \gg 1$.}
\label{approx3}
\end{equation}

If $K \ll 1$ then $E$ is close to $1$ and it is convenient to define a parameter $\delta\ll 1$ by $E = 1 -\delta^2$. Thus in \eqref{mathieu2} the growth rate $\cos y - E$ is positive only in a small region of size $\delta$ centered on $y=0$, and within this oasis the equation is
\begin{equation}
KQ_{yy} + \left( \delta^2 - y^2/2 \right) Q = O(\delta^4) \, .
\label{shm}
\end{equation}
The solution of this quantum oscillator problem is standard: $Q\approx  \exp(-y^2/4 \delta^2)$ and $K\approx 2 \delta^2$, or
\begin{equation}
K \approx 2(1-E)^2\, ,  \quad \text{if $K \ll 1$.}
\label{approx4}
\end{equation}

The two approximations in \eqref{approx3} and \eqref{approx4} can be combined into a single expression 
\begin{equation}
K \approx \frac{2(1+2E)(1-E)^2}{4E- E^2}\, .
\label{approx5}
\end{equation}
The above is asymptotically exact as $E\to 0$ and as $E\to 1$, and  interpolates the function $K(E)$ over the range $0<E<1$ with maximum error of about $4.5\%$ at $E\approx .67$.

The first application of \eqref{approx5} is to the eigenproblem obtained by setting $s_*=0$ in \eqref{zeroUeigen1}. Making the identifications $K=\kappa_*/(1-\Gamma)$ and $E = -\Gamma/(\Gamma-1)$ we obtain \eqref{mathieu2}. Then rewriting  \eqref{approx5} in terms of $\kappa_*$ and $|\Gamma | =-\Gamma$ we obtain the simple approximation \eqref{kluge1}. 
Figure \ref{zeroUFig}(b) compares this approximation with the numerical solution obtained by adapting program 21 of Trefethen \cite{LNTrefethen2000}.

\section{Optimization \label{optApp}}

We continue from \eqref{BP1} filling in some algebraic details.  Dropping the square containing $P$ from the right-hand side yields:
 \begin{equation}
\PP \leq   \alpha \B + \frac{ \la(\gamma + \beta \gamma - \alpha)^2 \ra}{4 \beta \eta} \, .
\label{opt1}
\end{equation}
Minimizing the right hand side of \eqref{opt1} over $\beta$ we find the optimal value of $\beta$: 
\begin{equation}
\beta_* = \sqrt{\frac{\la(\gamma - \alpha)^2\ra}{\la \gamma^2\ra}}  > 0\, .
\end{equation}
Plugging $\beta_*$ into \eqref{opt1} yields
\begin{equation}
\PP \leq \alpha \B  + \frac{\la \gamma(\gamma -\alpha)\ra + \sqrt{\la \gamma^2\ra\la(\gamma - \alpha)^2 \ra}}{2 \eta} \, .
 \label{BP2}
 \end{equation}
 (Note that $\beta_*$ is positive and so our earlier assumption that $\beta > 0$ is valid and we have found an upper bound.) Next we minimize the right-hand side of \eqref{BP2} over $\alpha$. After some work, we find that the optimal $\alpha_*$ satisfies
 \begin{equation}
 \frac{\la \gamma \ra - \alpha_*}{\sqrt{ \la (\gamma - \alpha_*)^2 \ra }} =  \A\, ,
  \label{BP3}
 \end{equation}
 where
 \begin{equation}
 \A \equiv \frac{2 \eta}{\sqrt{ \la \gamma^2 \ra }}\left(\B - \frac{\la \gamma \ra}{2 \eta} \right) \, .
  \label{BP3.1}
 \end{equation}
The inequality \eqref{summary} ensures that $0\leq \A \leq 1$. 
 
 Solving \eqref{BP3}, and taking the branch consistent with $\A \geq 0$, we find that
 \begin{equation}
 \alpha_* = \la \gamma \ra - \frac{\A \sigma }{\sqrt{1 - \A^2}} \, , 
  \label{BP4}
 \end{equation}
 where
 \begin{equation}
 \sigma \equiv \sqrt{\la \gamma^2\ra  - \la \gamma \ra^2}\, .
  \label{BP5}
 \end{equation}
 Substituting \eqref{BP4} into \eqref{BP2} we obtain
 \begin{equation}
 \PP \leq \la \gamma \ra \B + \frac{\sigma^2}{2 \eta} + \sigma \sqrt{\frac{\la \gamma^2 \ra}{4 \eta^2} -\left(\B  - \frac{\la \gamma\ra}{2 \eta}\right)^2}\, .
 \label{BPupper1}
 \end{equation}
Repeating the above procedure with $\beta < 0$ gives the other half of the bounding ellipse.

\bibliography{APSabbrevs,birchRefs}

\end{document}